# On a joint quantized and mechanical description for the Chernov-Lüders macroband of localized deformation

A.A. Reshetnyak[1,a)], E.V. Shilko[1,2b)], Yu.P. Sharkeev[1,3c)]

[1]*Institute of Strength Physics and Materials Science SB RAS, Akademicheskii Ave. 2/4, Tomsk, 634055, Russia*
[2]*National Research Tomsk State University, Novosobornaya, Sq. 1, Tomsk, 6340050, Russia*
[3]*National Research Tomsk Polytechnic University, Lenin Ave. 10, Tomsk, 634011, Russia*

[a)]Corresponding author: reshet@ispms.tsc.ru
[b)]Shilko@ ispms.tsc.ru, [c)]Sharkeev@ispms.tsc.ru,

**Abstract.** We suggest a quantum procedure, based on our recent statistical theory of flow stress in polycrystalline materials under quasi-static plastic deformations, with the intention to approach a theoretical description of the Chernov–Lüders shear macroband of localized deformation, exhibited by some Fe-containing materials with a second phase beyond the yield-strength point on the stress-strain curve $\sigma=\sigma(\varepsilon)$. The procedure makes substantial use of a quasi-particle interpretation for the minimal portion of mechanical energy in a given single-mode polycrystalline aggregate that is necessary for the thermal-fluctuation mechanism to create a 0D-defect nanopore as the initial zone of a localized deformation under external loading. Using a quasi-particle description, we obtain analytic expressions both for the scalar density of dislocations, given the size of grains, the temperature, the most probable sliding system, and for the dependence $\sigma=\sigma(\varepsilon)$ itself. A two-level system, which characterizes the mechanism of absorption and emission of such quasi-particles (dislocons) by the crystal lattice of any grain under quasi-static loading provides an effective physical description for the emergence and propagation of the Chernov–Lüders shear macroband. An enhancement of acoustic emission observed in experiments and accompanied by the macroband phenomenon justifies the interpretation of a dislocon as a composite short-lived particle consisting of acoustic phonons. A more realistic three-level system within a two-phase model with third (with dispersion particles) phase presence for actual polycrystalline samples is also proposed.

## INTRODUCTION

One of the principal trends in materials science is a search for controlling the internal defect substructure of crystallites with the intention to provide the best strength and plastic characteristics of polycrystalline (PC) materials. Optimizing these properties demands some new technologies, including the methods of severe plastic deformation combined with recrystallization annealing, the vapor deposition method, etc [1]. Such technologies allow for ample variations in the linear size *d* and orientation of microstructure elements for a given material: from meso-polycrystalline and coarse-grained (CG, 10−1000 μm) to fine-grained (FG, 2−10 μm), ultrafine-grained (UFG, 0.5−2.0 μm), sub-microcrystalline (SMC, 100−500 nm), and down to nano-crystalline (NC, <100 nm) samples. Experimental search for the physico-mechanical properties of PC materials (microhardness *H*, yield strength $\sigma_y$, ultimate stress $\sigma_S$, and strain hardening coefficient θ) has revealed such features of the hardening mechanism in the course of transition to the UFG, SMC and NC stages that are manifested by deviations from an empirical Hall–Petch (HP) relation [2], $\sigma_y(d) = \sigma_0 + kd^{-1/2}$, with $\sigma_0$ and *k* being the frictional stress (caused by dislocations as they move inside the grains) and the HP coefficient, respectively. This experimental research was continued by R. Armstrong, H. Conrad, U.F. Kocks, G. Langford, A.W. Thompson, J.G. Sevillano, S.A. Firstov, B.A. Movchan, V.I. Trefilov, Yu.Ya. Podrezov, V.V. Rybin, V.A. Likhachev, R.Z. Valiev, V.E. Panin, E.V. Kozlov, N.A. Koneva, A.D. Korotaev and A.N. Tyumentsev. However, no theoretical justification for the HP law was presented to explain the existence and behavior of maxima in $\sigma_y$, $\sigma_S$ or reveal any analytic dependence $\sigma = \sigma(\varepsilon,d)$ under plastic deformations (PD) outside the notions of continuous media mechanics. In our recent works [3, 4 ,5, 6], we have suggested a theoretical model of mechanical energy distribution in every crystallite of a single-mode isotropic PC material with respect to quantized quasi-stationary zones (levels of a finite width in the energy spectrum of a crystal lattice, CL) under quasi-static PDs. The energy spectrum of each crystallite was, in first approximation, was chosen to be equidistant, thereby implementing the most probable assembly of dislocations with a one and the same Burgers vector, *b*, starting from the zero-level energy of a crystallite without defects, $E_0$, and continued to the level of a maximal number of atoms in a (full) dislocation situated on a dislocation axis, $E_N$, $N(d)=[d/b]$. The expressions for statistically determined scalar density of dislocations (SDD) $\rho(b_\varepsilon,d,T,\varepsilon)$, flow stress (FS), including, $\sigma_y$, for single-mode PC aggregates under quasi-static

PD with velocities: $\dot{\varepsilon} \sim 10^{-5} - 10^{-3} \, s^{-1}$, depending on the average diameter $d$ of a crystallites (grains), as well as on a value of a (weak) grain-boundary un-hardening phase have been obtained in [3], [4]. In [4] it was suggested a quasiparticle interpretation for the portion (quantum) of energy, being equal to the unit dislocation energy, $E_d^{Le} = \frac{1}{2}Gb_\varepsilon^3$ with effective vector $b_\varepsilon = b(1 + \varepsilon)$, to be necessary for generation of the dislocation (within scenario [3] of a dislocation origin under PD from sequences of 0$D$-defects, in turn, thermal-fluctuation causes beyond the elastic limit under loading in the zones of localized plasticity (ZLP))[1]. It may be interpreted as a composite quasi-particle (made up by bounded acoustic phonons), provisionally called a *dislocon*, with the energy related to the frequency $\omega_\varepsilon$ according to the corpuscular-wave hypothesis of L. de Broglie, $\hbar\omega_\varepsilon = \frac{1}{2}Gb_\varepsilon^3$, and released (as the atomic bonds are broken up by a PD) to be possibly absorbed later on by a CL grain located close to the atom, which implies a local restoration of the CL translation symmetry or a local growth of the temperature.

The quasi-particle description of quasi-static deformations in PC materials allows one to solve numerous problems earlier regarded only in the mechanical approach. Thus, one can derive quantized expressions for SDD $\rho(b_\varepsilon,d,T)$ and then use the Taylor strain-hardening mechanism for FS to obtain σ(ε), including $\sigma_y$. This also provides an elegant physical approach to the emergence and propagation of the Chernov–Lüders (ChL) shear macroband on the basis of so-called two- and three-level systems. The paper is devoted to these three problems examined in the respective sections. A grain is understood as a crystallite with a small initial (prior to PD) density of dislocations.

## TWO-LEVEL SYSTEM FOR A DERIVATION OF THE SCALAR DENSITY OF DISLOCATIONS AND STRESS-STRAIN DEPENDENCE.

Let us introduce a concept of dislocons emission and absorption by using a simple *two-level system* for the energy spectrum of an arbitrary crystallite (grain) of single-mode PC aggregate under a quasi-static PD for given values of temperature and strain ε. $E_m(\varepsilon)$ and $E_n(\varepsilon)$ ($E_m<E_n$) for 0<m<n≤ $N$. In the course of an elementary PD act (defect origin) under constant external tensile loading, the crystallite at the instant $t = \varepsilon/\dot{\varepsilon}$ passes from a state of energy $E_m(\varepsilon)$ to a state of energy $E_n(\varepsilon)$, which is accompanied by the absorption of a dislocon of energy $\hbar\omega_{\varepsilon|nm}$:

$$\hbar\omega_{\varepsilon|nm} = \frac{(n-m)}{2}Gb_\varepsilon^3 = \mathrm{E}_n(\varepsilon) - \mathrm{E}_m(\varepsilon). \tag{1}$$

Given this, let us consider two kinds of energy transition from $E_m(\varepsilon)$ to $E_n(\varepsilon)$ at the thermodynamic equilibrium of a grain for a certain strain value ε between some PD acts of minimal duration $\Delta t_0 \sim 10^{-1}$s $\gg \tau = a/v_{зв} \sim 10^{-12}$ s at $\dot{\varepsilon}$=10$^{-5}$ s$^{-1}$ and with a relaxation time for the grain atoms at the new ground states[2].

In the first kind of transition (defect origin), the crystallite absorbs a quantum of energy being equal to the dislocon energy, $E_m(\varepsilon) \rightarrow E_n(\varepsilon)$.

In the second kind of transition, $E_n(\varepsilon) \rightarrow E_m(\varepsilon)$, characterized by local CL restoration (defect disappearance) or by transition to a defect state of lower energy, emission occurs of an energy quantum equal to the dislocon energy. In PC samples, there are many such crystallites. We distinguish (according to A. Einstein) between *spontaneous* and *induced emission* of such energy portions, as well as *induced absorption* at thermodynamic equilibrium with a given value of ε.

Induced absorption (defect emergence) takes place only under the influence of external loading with a given ε. Spontaneous emission occurs randomly, accompanied by the creation, in various crystallite parts, of dislocons with different properties (non-coherent waves). Induced emission at transitions $E_n(\varepsilon) \rightarrow E_m(\varepsilon)$ is possible if a localized portion of energy (dislocon A) is carried to one of the CL atoms (nodes), which is an excited "defect" state leading to an emission (as this quantum of energy interacts with the crystallite) of a new dislocon (*B*), with the same properties as those of dislocon A (coherent waves).

The counting of an average number of transitions $dN_{nm}^\varepsilon$ from a state of energy $E_n(\varepsilon)$ to a state of energy $E_m(\varepsilon)$ during a time interval $dt < \Delta t_0$ in an arbitrary crystallite of the PC sample under spontaneous emission shows that $dN_{nm}^\varepsilon$ is proportional to the transition probability $P_{nm}^\varepsilon$ and to the number of grain atoms (defects) $N_n^\varepsilon$ that characterizes the upper energy level $E_n(\varepsilon)$ per unit volume, which implies the representation

$$dN_{nm}^\varepsilon = P_{nm}^\varepsilon N_n^\varepsilon \, dt. \tag{2}$$

For an induced (stimulated) dislocon emission during the same $dt$ inside $[t, t + dt]$, we find that the average number of such transitions, $dN_{nm}^{\varepsilon|ind}$, is proportional to the probability of induced transitions, $P_{nm}^{\varepsilon|ind}$, the quantity of SDD, $\rho(b_\varepsilon,d,T,\varepsilon)$, in a given grain with $b_\varepsilon = b_\varepsilon(\omega_{\varepsilon|nm})$ [because of the more number of dislocations, thus the ones of dislocons per unit volume, the more $dN_{nm}^{\varepsilon|ind}$], as well as to the population $N_n^\varepsilon$ of grain atoms (defects) with $\mathrm{E}_n(\varepsilon)$

$$dN_{nm}^{\varepsilon|ind} = P_{nm}^{\varepsilon|ind} N_n^\varepsilon \, \rho(b_\varepsilon, d, T, \varepsilon) \, dt. \tag{3}$$

Finally, at an induced absorption of the quantized portion of mechanical energy by a given crystallite, the average number of transitions $E_m(\varepsilon) \rightarrow E_n(\varepsilon)$ per unit 1 m$^3$, denoted by $dN_{mn}^{\varepsilon|ind}$, at the same interval of time $[t, t + dt]$ as in (2), (3), albeit with $(P_{nm}^{\varepsilon|ind}, N_n^\varepsilon) \to (P_{mn}^{\varepsilon|ind}, N_m^\varepsilon)$ and with symmetric probabilities $P_{nm}^{\varepsilon|ind} = P_{mn}^{\varepsilon|ind}$, is determined entirely by the choice of a PC material:

$$dN_{mn}^{\varepsilon|ind} = P_{mn}^{\varepsilon|ind} N_m^\varepsilon \rho(b_\varepsilon, d, T, \varepsilon) \, dt. \tag{4}$$

---

[1] The value of $E_d^{L_e}$ is commensurate with the activation energy of an atom in a material during the diffusion process [3, 5].

[2] A PD process, generally non-equilibrium one, is presented [3, 5] as a sequence of equilibrium PD processes, abruptly changing from one PD act to another.

The principle of detailed equilibrium at a given ε implies the following relation for a grain:

$$dN_{nm}^\varepsilon + dN_{nm}^{\varepsilon|ind} = dN_{mn}^{\varepsilon|ind} \Rightarrow \frac{N_n^\varepsilon}{N_m^\varepsilon} = \frac{P_{mn}^{\varepsilon|ind} \rho(b_\varepsilon,d,T,\varepsilon)}{P_{nm}^\varepsilon + P_{nm}^{\varepsilon|ind} \rho(b_\varepsilon,d,T,\varepsilon)} \tag{5}$$

At thermodynamic equilibrium (in the equilibrium state segment, according to Footnote 2) for a given ε, crystallites occupy the energy levels $E_i(\varepsilon)$, $i = n, m$, according to the Boltzmann distribution (which, in turn, may be obtained from the Large Numbers Law) $W_i = C \exp\left(-\frac{E_i(\varepsilon)}{NkT}\right)$. From $\frac{N_n^\varepsilon}{N_m^\varepsilon} = \frac{W_n}{W_m}$ and (1), (5), with account taken of $\hbar\omega_{\varepsilon|n,n-1} = \frac{1}{2}Gb_\varepsilon^3$, it follows that:

$$\rho(b_\varepsilon, d, T, \varepsilon) = P_{nm}^\varepsilon / \left\{P_{nm}^{\varepsilon|ind}\left(\exp\left(\frac{(n-m)Gb_\varepsilon^3}{2NkT}\right) - 1\right)\right\} \text{ and } \lim_{N=\frac{d}{b_\varepsilon}\gg b} \rho(b_\varepsilon, d, T, \varepsilon) = \frac{P_{nm}^\varepsilon}{P_{nm}^{\varepsilon|ind}} \frac{2dkT}{Gb_\varepsilon^3(n-m)}. \tag{6}$$

Using the explicit form of $\rho(b_\varepsilon, d, T, \varepsilon)$ in the CG limit [3, 4], $\rho(b_\varepsilon, d, T, \varepsilon)\big|_{\frac{d}{b}\gg b} = \frac{12}{\pi} \frac{\varepsilon m_0}{b_\varepsilon d(1+\varepsilon)}$, with allowance for a specific distribution of crystallites in a PC material with a certain texture to be incorporated analytically into the general SDD $\bar\rho(b_\varepsilon, d, T, \varepsilon)$ for a real-valued average texture factor $\bar K$, we have

$$\bar\rho(b_\varepsilon, d, T, \varepsilon)\big|_{\frac{d}{b}\gg b} = \rho(b_\varepsilon, d, T, \varepsilon)\bar K = \bar K \frac{12}{\pi} \frac{\varepsilon m_0}{b_\varepsilon d(1+\varepsilon)} \quad \text{for } 0 < \bar K \le 1, \tag{7}$$

where $m_0$ is a *polyhedral parameter* [3,5] taking into account the number of crystallographic planes contributing to a crystallite deformation due to its polyhedral character. . For a cubic CL, an isotropic distribution of crystallites in a PC sample implies that the crystallographic slip planes relative to the loading z-axis are situated inside the angle $\left[-\frac{\pi}{4}, \frac{\pi}{4}\right]$, so that averaging, with respect to all directions, of the texture factor $K = K(x, y, z)$ leads to $\bar K = 1/\sqrt{2}$ [3,5]. From (6), (7) it follows that: $P_{nm}^\varepsilon / P_{nm}^{\varepsilon|ind} = \varepsilon m_0 (n-m) \frac{Gb^2}{2kTd^2} \frac{6\sqrt{2}}{\pi}$, and therefore the SDD in the whole single-mode PC material with the grains distributed isotropically and with an extreme grain size $d_\rho$ for an SDD maximum $\bar\rho(b_\varepsilon, d, T, \varepsilon)_m$ is determined using the equation: $\partial\rho(b_\varepsilon, d, T, \varepsilon)/\partial d = 0$:

$$\bar\rho(b_\varepsilon, d, T, \varepsilon) = M(0) \frac{6\sqrt{2}}{\pi} \frac{m_0}{d^2} \varepsilon \left(e^{M(\varepsilon)b/d} - 1\right)^{-1} + o(\varepsilon^2), \quad M(\varepsilon) = Gb_\varepsilon^3/2kT, \quad d_\rho(\varepsilon, T) = b\frac{Gb^3(1+\varepsilon)^3}{2 \cdot 1{,}59363 \cdot kT}. \tag{8}$$

Note, that $d_\rho(\varepsilon, T)$ coincides with the extreme size at a maximum of $\sigma(\varepsilon)$: $d_\rho(\varepsilon, T) = d_0(\varepsilon, T)$. The analytical dependence $\sigma(\varepsilon)$, in particular, $\sigma_y = \sigma(0.002)$, is given by the Taylor strain hardening mechanism [3], $\tau = \tau_f + \alpha G b \sqrt{\rho}$, $\sigma(\varepsilon) = m\tau$, $m = 3.05$ with a dislocation interaction constant α and without a grain boundary phase:

$$\sigma(\varepsilon) = \sigma_0(\varepsilon) + \alpha m \frac{Gb}{d} \sqrt{\frac{6\sqrt{2}}{\pi} m_0 \varepsilon M(0)} \left(e^{M(\varepsilon)\left[\frac{b}{d}\right]} - 1\right)^{-\frac{1}{2}}. \tag{9}$$

The graphic dependence of SDD (with initial for $\varepsilon = 0$ SDD $\rho_0$: $\rho_0 \ll \bar\rho$) as a function of the average diameter $d$ and its inverse square root $d^{-1/2}$ at ε=0.002, ε=0.01, T=300K for single-mode one-phase PC aggregates of α- Fe (BCC CL); Cu, Al, Ni (FCC CL); α-Ti, Zr (HCP CL) is shown in Fig. 1, with account taken of Table 1. The latter contains the values of $G$, the lattice constant $a$ [10], the Burgers vectors of the least possible length $b$, the respective most probable sliding systems (see Table 2 [4]), the interaction constant for dislocations $\alpha$ [6,7], the calculated values for dislocons (smallest unit dislocations $E_d^{L_e}$), and the extreme grain size $d_\rho(0.002, 300)$ for the maxima of $\bar\rho(b_\varepsilon, d, T, 0.002)_m$, $\bar\rho(b_\varepsilon, d, T, 0.01)_m$, in accordance with (5) [3]. The values of $m_0$ are determined in the CG limit of the HP law: $\sigma(\varepsilon)\big|_{d \gg b} = \sigma_0(\varepsilon) + k(\varepsilon)d^{-1/2} \Rightarrow k(\varepsilon)\sqrt{(1+\varepsilon)^3/\varepsilon} = \alpha m G \sqrt{\frac{6\sqrt{2}}{\pi} m_0} b$, where $m_0 = m_0(k^2(\varepsilon))$.

**TABLE 1**: The values of $\bar\rho_m$, $E_d^{L_e}$, $k$, $m_0$, α in BCC, FCC and HCP polycrystalline metal samples with $d_0$, $b$, $G$ obtained using the data of [1,4,6] at $\varepsilon = 0{,}002$, T=300K for α-Fe; Cu; Al, Ni, α-Ti, Zr single-mode PC aggregates with $\rho_0$=0.

| Type of CL | BCC | FCC | | | HCP | |
|---|---|---|---|---|---|---|
| Material | α-Fe | Cu | Al | Ni | α-Ti | Zr |
| $b$, nm | $\frac{\sqrt{3}}{2}a$ =0.248 | $a/\sqrt{2}$ =0.256 | =0.286 | $a/\sqrt{2}$ =0.249 | $a$=0.295 | $a$=0.323 |
| $G$, GPa | 82.5 | 44 | 26.5 | 76 | 41.4 | 34 |
| $k(0.002)$, MPa $m^{1/2}$ | 0.55-0.65 ($10^{-5}-10^{-3}m$) | 0.25 ($10^{-4}-10^{-3}m$) | 0.15 ($10^{-4}-10^{-3}m$) | 0.28 ($10^{-5}-10^{-3}m$) | 0.38-0.43 ($10^{-5}-10^{-3}m$) | 0.26 ($10^{-5}-10^{-3}m$) |
| α | 0.30 | 0.38 | 0.27 | 0.35 | 0.97 | - |
| $E_d^{L_e} = \frac{1}{2}Gb^3$, eV | 3.93 | 1.28 | 1.96 | 3.72 | 3.33 | 3.57 |
| $m_0$ | 40.7-56.8 | 17.8 | 31.3 | 9.1 | 6.2-8.0 | 14.8 |
| $d_\rho = d_0$, nm | 23.6 | 14.4 | 13.6 | 22.6 | 23.8 | 28.0 |
| $\rho_m(0.002)$ $*10^{-15}$, m$^{-2}$ | 15.1 | 10.5 | 17.7 | 3.6 | 2.3 | 3.5 |
| $\rho_m(0.01)$ $*10^{-15}$, m$^{-2}$ | 71.2 | 50.2 | 84.2 | 17.1 | 11.1 | 16.9 |

The values of $k(\varepsilon)$ at $\varepsilon = 0.002$ chosen, e.g. for α-Fe, Cu, Ni [6,7], Al [10], Zr, α-Ti [2,8,9] within the grain range enclosed in the frames and the value of $\alpha$ for Zr approximately equal to 0.5.

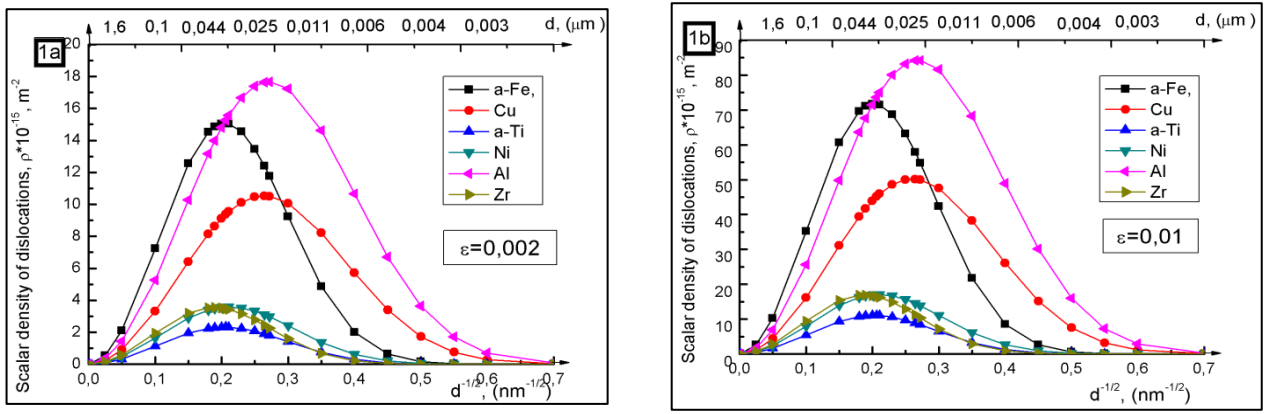

**FIGURE 1**: The graphic dependence for the scalar density of dislocations (8) $\bar{\rho} = \bar{\rho}(d^{-1/2})$ in single-mode 1-phase PC materials at $\varepsilon = 0.002$ in Fig. 1a, $\varepsilon = 0.01$ in Fig. 1b, T=300K with an additional upper scale and the size of grains $d$ measured in μm. The upper axis lies in the range (∞;0) with the inverse quadratic scale and the correspondence (100; 1.6; 0.1; 0.044; 0.025; 0.011; 0.006; 0.004; 0.003) μm ↔ (0.005; 0.015; 0.1; 0.15; 0.2; 0.3; 0.41; 0.5; 0.57) nm$^{-1/2}$ for the respective values on the lower axis. The least possible values of the parameter $m_0(k)$ for α-Fe, α-Ti for the maxima of $\sigma_y$ are calculated using the respective (see Table 1) extreme grain size values $d_\rho$.

## TWO- AND THREE-LEVEL SYSTEM QUANTIZED APPROACH TO CHERNOV-LÜDERS MACROBANDS DESCRIPTION

The derivation procedure for $\bar{\rho}(b_\varepsilon, d, T, \varepsilon)$ and $\sigma(\varepsilon)$ in (8), (9) opens a perspective to a physical description of the emergence and propagation of ChL shear macrobands for localized PDs in PC materials. This macroscopic phenomenon arises explicitly when a yield surface does exist for a given PC aggregate (normally with dispersion hardening) and is accompanied by a quasi-coherent emission (due to a non-uniform distribution of PC sample grains) of collective phonons (i.e., dislocons in our quasi-particle interpretation), which is manifested by a multiple amplification of acoustic emission in such samples (see Figs. 2, 3 [10] for the yield surface in low-alloy steel)

For this purpose, let us introduce a simple three-level system with an intermediate third level $E_k(\varepsilon)$ in addition to $E_n(\varepsilon)$, $E_n(\varepsilon)$ ($E_m(\varepsilon) \ll E_k(\varepsilon) < E_n(\varepsilon)$), $0 < m < k < n \leq N$, outside the equidistant spectrum approximation for a crystallite. The process of ChL macroband emergence may be described by the following sequence of quasi-static PDs.

First, on a yield surface (e.g, in the case of austenite and ferrite steels) under quasi-static tensile PDs in the entire set of crystallites, one observes, first of all, a multiple ZLP formation in which the inverse population of energy levels $N_n^\varepsilon$ ($N_n^\varepsilon > N_k^\varepsilon > N_m^\varepsilon$) takes place for a given $\varepsilon$ by means of induced localized absorption of energy by the crystallite. From the thermodynamic viewpoint, it is favorable for every crystallite to form a defect-dislocation substructure with the least possible energy for every crystallite ($N_n^\varepsilon < N_k^\varepsilon < N_m^\varepsilon$). Second, the lifetime of a defect with energy $E_n(\varepsilon)$ is small, and a spontaneous emission of dislocons makes the crystallite pass to a state with the lowest defect energy $E_k(\varepsilon)$ and the population $N_n^\varepsilon < N_k^\varepsilon$, $N_k^\varepsilon > N_m^\varepsilon$. Third, the energy extracted in this process ($N_n^\varepsilon - N_k^\varepsilon)(E_n(\varepsilon) - E_k(\varepsilon))$ is used for a further expansion of the ZLP in size. Fourth, the procedure takes place (in a certain grain at the PC sample boundary as the flow stage is most pronounced and reaches its maximal inverse population, $N_k^\varepsilon > N_m^\varepsilon$) until a multiple transition emerges from a state of energy $E_k(\varepsilon)$ to a stage of energy $E_m(\varepsilon)$ with a multiple extraction of energy, with account taken of the induced emission of quasi-coherent dislocons. Fifth, the extracted energy of dislocons is carried along the crystallite atoms from one ZLP to another, and so on, leading to their accommodation and repeating the process of dislocon energy extraction until reaching the opposite grain boundary. Further, the process goes on according to the above procedure in the neighboring (already prepared) crystallites with an inverse population of crystallite energy levels in the ZLP, and so on.

As a result, a ChL macroband visually originates from a sample boundary near the point of attachment to the opposite boundary at a certain (non-)constant angle (near π/4), due to additional shear deformations of every grain involved, and thereby a propagation front arises as an autowave with respect to the loading axis directed along the entire sample. One can generate multiple ChL macrobands when the above conditions for an inverse population of crystallite energy levels can be repeated.

Mechanically, the process is provided by so-called martensite transformations occurring in the ZLP under the action of a high internal local stress, which ensures a direct transition from the FCC γ-phase to the BCC α(α')-phase in the CL, which is then after some shear followed by an opposite transition, while producing twinning-type defects with allowance for combinations of partial Shockley dislocations of lowest energies and already decreased local values in the crystallite (close to the sample boundary), accompanied by a ChL macroband emergence and followed by subsequent crystallites, which was shown in Figs. 2–4 [11] for nearly ferrite steel.

## SUMMARY

Our quantized approach based on a quasi-particle interpretation for a portion of energy being equal to a unit dislocation energy within a recently proposed statistical theory of flow stress in PC materials under quasi-static plastic deformations allows one (using two-level systems with an equidistant energy spectrum in an arbitrary crystallite (grain) of a single-mode PC aggregate) to obtain the analytic dependence for scalar density of dislocations (8) with a set of few natural parameters and the flow stress (9) within the Taylor strain-hardening mechanism in the entire range of grain size values for single-mode PC aggregates (without a weak phase). The suggested concept of (*spontaneous* and *induced*) emission and (*induced*) absorption of disclons by the CL of a PC sample, as the defect structure, respectively, decreases and increases under quasi-static PDs, proves to be crucial for an explanation of microphysical deformation processes and for a derivation of the above integral mechanical characteristics. To illustrate our general results, a graphic dependence of SDD as functions of an average inverse square root of the diameter, $d^{-1/2}$, at ε=0.002 and ε=0.01 for T=300K, in single-mode one-phase PC aggregates of α-Fe (BCC CL), Cu, Al, Ni (FCC CL), α-Ti, Zr (HCP CL) is presented in a good agreement with experimental data.

An extension of two-level system to three-level one, with an *intermediate* third level $E_k(\varepsilon)$, in addition to $E_m(\varepsilon)$, $E_n(\varepsilon)$ ($E_m(\varepsilon) \ll E_k(\varepsilon) < E_n(\varepsilon)$), 0<m<k<n≤ N, besides the levels used in the equidistant spectrum approximation for a crystallite [3,4,5,6], allows one to efficiently describe the process of emergence and propagation of a ChL macroband in accordance with its standard mechanical description [11], thereby revealing the nature of acoustic emission growth [10] in austenite and ferrite steel samples (with third-phase particles C, Mn corresponding to dispersion hardening and non-equidistant grain energy spectra) under quasi-static tensile PDs on yield surfaces.

The model proposed to explain the ChL macroband physics can be used directly for a description of the related Portevin–Le Chatelier effect.

## ACKNOWLEDGMENTS

The authors are grateful to the research fellows of ISPMS SB RAS for useful discussions. The work is supported by the Program of Fundamental Research under the Russian Academy of Sciences, 2013–2020, (direction III.23).